%% file: main.tex
\documentclass[sigconf]{acmart}

\acmSubmissionID{526}

\usepackage{subcaption}
\usepackage{array}
\newcolumntype{P}[1]{>{\centering\arraybackslash}p{#1}}
\usepackage{multirow}

\graphicspath{{figures/}{pictures/}{images/}{./}} 

\usepackage[ruled]{algorithm2e} 

\SetAlCapHSkip{0pt}
\SetKwInput{KwInput}{Input}                
\SetKwInput{KwOutput}{Output} 

\copyrightyear{2022}
\acmYear{2022}
\setcopyright{acmcopyright}\acmConference[SIGGRAPH '22 Conference Proceedings]{Special Interest Group on Computer Graphics and Interactive Techniques Conference Proceedings}{August 7--11, 2022}{Vancouver, BC, Canada}
\acmBooktitle{Special Interest Group on Computer Graphics and Interactive Techniques Conference Proceedings (SIGGRAPH '22 Conference Proceedings), August 7--11, 2022, Vancouver, BC, Canada}
\acmPrice{15.00}
\acmDOI{10.1145/3528233.3530731}
\acmISBN{978-1-4503-9337-9/22/08}

\begin{document}
\title{GWA: A Large High-Quality Acoustic Dataset for Audio Processing}

\author{Zhenyu Tang}
\email{zhy@umd.edu}
\author{Rohith Aralikatti}
\email{rohithca@umd.edu}
\affiliation{%
 \institution{University of Maryland}
 \department{Department of Computer Science}
 \city{College Park}
 \state{MD}
 \postcode{20742}
 \country{USA}
}

\author{Anton Ratnarajah}
\email{jeran@umd.edu}
\affiliation{%
 \department{Department of Electrical and Computer Engineering}
 \institution{University of Maryland}
 \city{College Park}
 \state{MD}
 \postcode{20742}
 \country{USA}
}

\author{Dinesh Manocha}
\email{dmanocha@umd.edu}
\affiliation{%
 \department{Department of Computer Science, Department of Electrical and Computer Engineering}
 \institution{University of Maryland}
 \city{College Park}
 \state{MD}
 \postcode{20742}
 \country{USA}
}

\renewcommand\shortauthors{Tang, Z. et al}

\begin{abstract}
We present the Geometric-Wave Acoustic (GWA) dataset, a large-scale audio dataset of about 2 million synthetic room impulse responses (IRs) and their corresponding detailed geometric and simulation configurations. Our dataset samples acoustic environments from over 6.8K high-quality diverse and professionally designed houses represented as semantically labeled 3D meshes. We also present a novel real-world acoustic materials assignment scheme based on semantic matching that uses a sentence transformer model. We compute high-quality impulse responses corresponding to accurate low-frequency and high-frequency wave effects by automatically calibrating geometric acoustic ray-tracing with a finite-difference time-domain wave solver. We demonstrate the higher accuracy of our IRs by comparing with recorded IRs from complex real-world environments.  Moreover, we highlight the benefits of GWA on audio deep learning tasks such as automated speech recognition, speech enhancement, and speech separation. This dataset is the first data with accurate wave acoustic simulations in complex scenes. Codes and data are available at \url{https://gamma.umd.edu/pro/sound/gwa}.

\end{abstract}

%
%

%
%


\begin{teaserfigure} \centering \includegraphics[width=0.95\linewidth,trim={0 0.5cm 0 0},clip]{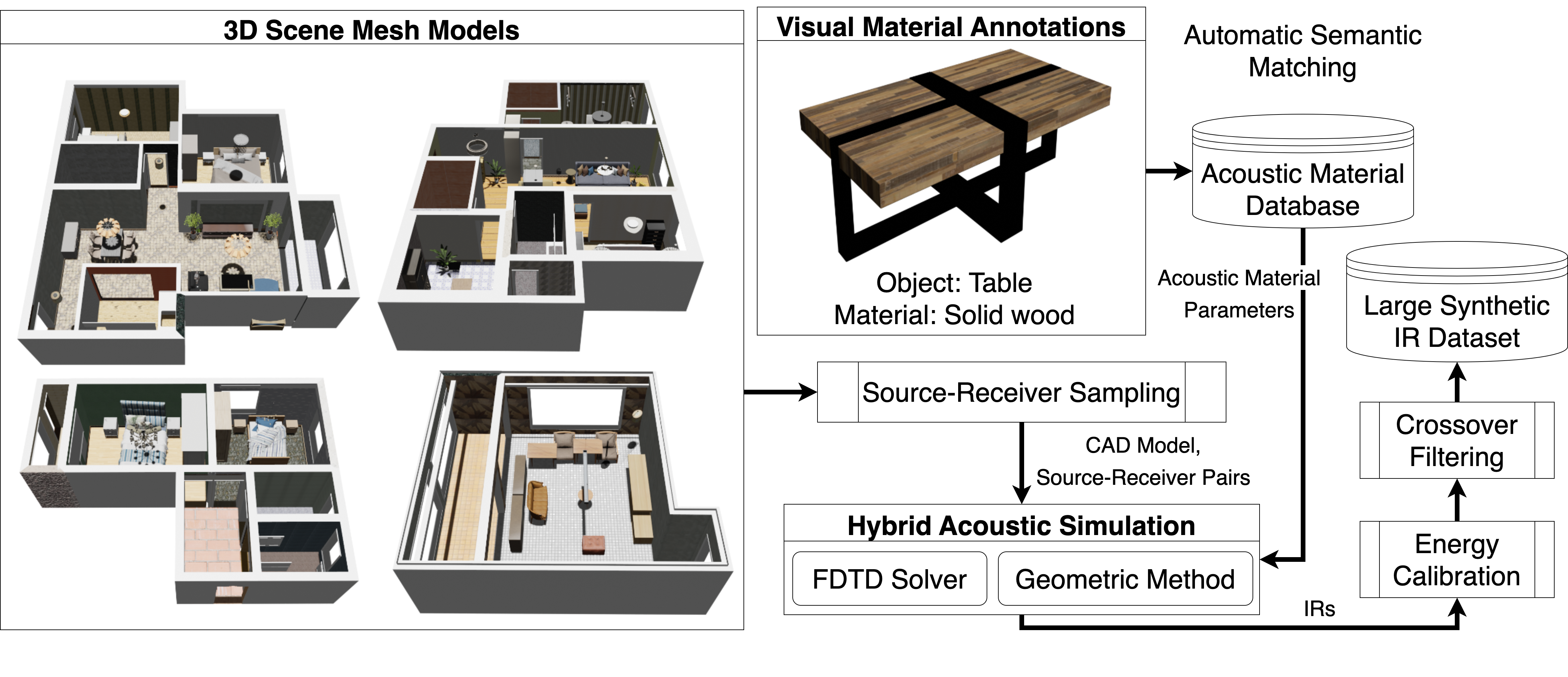}
\caption{Our IR data generation pipeline starts from a 3D model of a complex scene and its visual material annotations (unstructured texts). We sample multiple collision-free source and receiver locations in the scene. We use a novel scheme to automatically assign acoustic material parameters by semantic matching from a large acoustic database. Our hybrid acoustic simulator  generates accurate impulse responses (IRs), which become part of the large synthetic IR dataset after post-processing. }
\label{fig:overview}
\end{teaserfigure}

\maketitle
\input{1-intro.tex}

\input{3-dataset.tex}

\input{4-evaluation.tex}

\input{5-application.tex}

\input{6-conclusion.tex}

\begin{acks}
The authors acknowledge the University of Maryland supercomputing resources (http://hpcc.umd.edu) made available for conducting the research reported in this paper and UMIACS for providing additional computing and storage resources. The authors also thank Dr.\ Brian Hamilton for answering questions about the \emph{PFFDTD} software. This work is supported in part by ARO grants W911NF-18-1-0313, W911NF-21-1-0026, NSF grant \#1910940 and Intel. 
\end{acks}

\bibliographystyle{ACM-Reference-Format}
\bibliography{sound}

\end{document}

%% file: 1-intro.tex
\section{Introduction}
\label{sec:intro}
Audio signals corresponding to music, speech, and non-verbal sounds in the real world encode rich information regarding the surrounding environment.  Many digital signal processing algorithms and audio deep learning techniques have been proposed to extract information from audio signals. These methods are widely used for different applications such as music information retrieval, automated speech recognition, sound separation and localization, sound synthesis and rendering, etc.

Many audio processing tasks have seen rapid progress in recent years due to advances in deep learning and the accumulation of large-scale audio or speech datasets. Not only are these techniques widely used for speech processing, but also acoustic scene understanding and reconstruction, generating plausible sound effects for interactive applications, audio synthesis for videos, etc. A key factor in the advancement of these methods is the development of audio datasets. There are many datasets for speech processing, including datasets with different settings and languages~\cite{DBLP:journals/corr/abs-1903-11269}, emotional speech~\cite{tits2019emotional}, speech source separation~\cite{drude2019sms}, sound source localization~\cite{wu2018indoor}, noise suppression~\cite{reddy2020interspeech}, background noise~\cite{reddy2019scalable}, music generation~\cite{DBLP:journals/corr/abs-1709-01620}, etc.

In this paper, we present a large, novel dataset corresponding to synthetic room impulse responses (IRs). In the broader field of acoustical analysis, the IR is defined as the time domain (i.e. time vs amplitude) response of a system to an impulsive stimulus~\cite{kuttruff2016room}. It is regarded as the {\em acoustical signature} of a system and contains information related to reverberant, signal-to-noise ratio, arrival time, energy of direct and indirect sound, or other data related to acoustic scene analysis. These IRs can be convolved with anechoic sound to generate artificial reverberation, which is widely used in music, gaming and VR applications.
The computation of accurate IRs and resulting datasets are used for the following areas:

\paragraph{Sound Propagation and Rendering} Sounds in nature are produced by vibrating objects and then propagate through a medium (e.g., air) before finally being heard by a listener. Humans can perceive these sound waves in the frequency range of $20$Hz to $20$KHz (human aural range). There is a large body of literature on modeling sound propagation in indoor scenes using geometric and wave-based methods~\cite{liu2020sound,krokstad1968,vorlander1989,funkhouser1998,raghuvanshi2009efficient,mehra2013wave,schissler2014high,allen2015aerophones}. Wave-based solvers are expensive for high-frequency sound propagation. As approximations, \citet{wang2018toward} use full wave simulation to model local sound synthesis followed by estimated far-field sound radiation, while \citet{chaitanya2020directional} pre-compute and encode static sound fields using psycho-acoustic parameters. Geometric methods, widely used in interactive applications, are accurate for higher frequencies. We need automatic software systems that can accurately compute IRs corresponding to human aural range and handle arbitrary 3D models. 

\paragraph{Deep Audio Synthesis for Videos} Video acquisition has become very common and easy. However, it is hard to add realistic audio that can be synchronized with animation in a video. Many deep learning methods have been proposed for such audio synthesis that utilize acoustic impulse responses for such applications~\cite{Li:2018:360audio,owens2016visually,zhou2018visual}.

\paragraph{Speech Processing using Deep Learning} IRs consist of many clues related to reproducing or understanding intelligible human speech. Synthetic datasets of IRs have been used in machine learning methods for automatic speech recognition~\cite{malik2021automatic,ko2017study,tang2020improving,ratnarajah2020ir}, sound source separation~\cite{aralikatti2021improving,jenrungrot2020cone} and sound source localization~\cite{DBLP:journals/corr/abs-2109-03465}.

\paragraph{Sound Simulation using Machine Learning} Many recent deep learning methods have been proposed for sound synthesis~\cite{hawley2020synthesis,ji2020comprehensive,jin2020deep}, scattering effect computation, and sound propagation~\cite{fan2020fast,tang2021learning,pulkki2019machine}.  Deep learning methods have also been used to compute material properties of a room and acoustic characteristics~\cite{schissler2017acoustic,tang2019scene}.

Other applications that have used acoustic datasets include  navigation~\cite{chen2020soundspaces}, floorplan reconstruction~\cite{purushwalkam2021audio} and depth estimation algorithms~\cite{gao2020visualechoes}.

There are some known datasets of recorded IRs from real-world scenes and synthetic IRs (see Table~\ref{tab:comparison}). The real-world datasets are limited in terms of number of IRs or the size and characteristics of the captured scenes. All prior synthetic IR datasets are generated using geometric simulators and do not accurately capture low-frequency wave effects. This limits their applications. 

\setlength\tabcolsep{2pt}
\begin{table*}[t!]
  \caption{Overview of some existing large IR datasets and their characteristics. In the ``Type'' column, ``Rec.'' means \emph{recorded} and ``Syn.'' means \emph{synthetic}. The real-world datasets capture the low-frequency (LF) and high-frequency (HF) wave effects in the recorded IRs.  Note that all prior synthetic datasets use geometric simulation methods and are accurate for higher frequencies only. In contrast, we use an accurate hybrid geometric-wave simulator on more diverse input data, corresponding to professionally designed 3D interior models with furniture, and generate accurate IRs corresponding to the entire human aural range (LF and HF). We highlight the benefits of our high-quality dataset for different audio and speech applications. }
  \centering
    \begin{tabular}{c|c|c|c|c|c|c|c}
    \hline
    \textbf{Dataset} & \textbf{Type} & \textbf{\#IRs}  & \textbf{\#Scenes} &\textbf{Scene Descriptions} & \textbf{Scene Types} & \textbf{Acoustic Material} & \textbf{Quality}\\
    \hline\hline
    BIU~\cite{hadad2014multichannel} & Rec. & 234  & 3 & Photos & Acoustic lab & Real-world &   LF, HF  \\\hline
    MeshRIR~\cite{koyama2021meshrir} & Rec. & 4.4K  & 2 & Room dimensions & Acoustic lab & Real-world &    LF, HF \\\hline
    BUT Reverb~\cite{szoke2019building} & Rec. & 1.3K  & 8 & Photos & Various sized rooms & Real-world &    LF, HF \\\hline
    S3A~\cite{ColemanP2020SRIR} & Rec. & 1.6K  & 5 & Room dimensions & Various sized rooms & Real-world &    LF, HF \\\hline
    dEchorate~\cite{di2021dechorate} & Rec. & 2K  & 11 & Room dimensions & Acoustic lab & Real-world &    LF, HF \\\hline \hline
    \citet{ko2017study} & Syn. & 60K  & 600 & Room dimensions & Empty shoebox rooms  & Uniform sampling &    HF \\\hline
    BIRD~\cite{grondin2020bird} & Syn. & 100K  & 100K   & Room dimensions & Empty shoebox rooms  & Uniform sampling & HF \\\hline
    SoundSpaces~\cite{chen2020soundspaces} & Syn. & 16M  & 101   & Annotated 3D model & Scanned indoor scenes  & Material database & HF \\\hline
    GWA (ours) & Syn. & 2M   & 18.9K  & Annotated 3D model & Professionally designed  & Material database & LF, HF \\
    \hline
    \end{tabular}
  \label{tab:comparison}%
\end{table*}%

\paragraph{Main Results} We present the first large acoustic dataset (GWA) of synthetically generated IRs that uses accurate wave acoustic simulations in complex scenes. Our approach is based on using a hybrid simulator that combines a wave-solver based on finite differences time domain (FDTD) method with geometric sound propagation based on path tracing. The resulting IRs are accurate over the human aural range. Moreover, we use a large database of more than $6.8$K professionally designed scenes with more than $18$K rooms with furniture that provide a diverse set of geometric models. We  present a novel and automatic scheme for semantic acoustic material assignment based on natural language processing techniques. We use a database of absorption coefficients of $2,042$ unique real-world materials and use a transformer network for semantic material selection. Currently, GWA consists of more than 2 million IRs. We can easily use our approach to generate more IRs by either changing the source and receiver positions or using different set of geometric models or materials.
The novel components of our work include:
\begin{itemize}
    \item Our dataset has more acoustic environments than real-world IR datasets by two orders of  magnitude.
    \item Our dataset has more diverse IRs with higher accuracy, as compared to prior synthetic IR datasets.
    \item The accuracy improvement of our hybrid method over prior methods is evaluated by comparing our IRs with recorded IRs of multiple real-world scenes. 
    \item We use our dataset to improve the performance of deep-learning speech processing algorithms, including automatic speech recognition, speech enhancement, and source separation, and observe significant performance improvement. 
\end{itemize}

%% file: 3-dataset.tex
\section{Dataset Creation}

A key issue in terms of the design and release of an acoustic dataset is the choice of underlying 3D geometric models.
Given the availability of interactive geometric acoustic simulation software packages, it is relatively simple to randomly sample a set of simple virtual shoebox-shaped rooms for source and listener positions and generate unlimited simulated IR data. 
However, the underlying issue is such IR data will not have the acoustic variety (e.g., room equalization, material diversity, wave effects, reverberation patterns, etc.) frequently observed in real-world datasets. 
We identify several criteria that are important in terms of creating a useful synthetic acoustic dataset: (1) a wide range of room configurations: the room space should include regular and irregular shapes as well as furniture placed in reasonable ways. Many prior datasets are limited to rectangular, shoebox-shaped or empty rooms (see Table 1); (2) meaningful acoustic materials: object surfaces should use physically plausible acoustic materials with varying absorption and scattering coefficients,  rather than randomly assigned frequency-dependent coefficients; (3) an accurate simulation method that accounts for various acoustic effects, including specular and diffuse reflections, occlusion, and low-frequency wave effects like diffraction. It is important to generate IRs corresponding to the human aural range for many speech processing and related applications.
In this section, we present our pipeline for developing a dataset that satisfies all these criteria. 
An overview of our pipeline is illustrated in Figure~\ref{fig:overview}. 

\subsection{Acoustic Environment Acquisition}
Acoustic simulation for 3D models requires that environment boundaries and object shapes be well defined. This requirement can often be fulfilled by 3D meshes.
Simple image-method simulations may only require a few room dimensions (i.e., length, width, and height) and have been used for speech applications, but these methods cannot handle complex 3D indoor scenes. 
Many techniques have been proposed in computer vision to reconstruct large-scale 3D environments using RGB-D input~\cite{choi2015robust}. Moreover,  they can be combined with 3D semantic segmentation~\cite{dai2018scancomplete} to recover category labels of objects in the scene. 
This facilitates the collection of indoor scene datasets. 
However, real-world 3D scans tend to suffer from measurement noise, resulting in incomplete/discontinuous surfaces in the reconstructed model that can be problematic for acoustic simulation algorithms. 
One alternative is to use professionally designed scenes of indoor scenes in the form of CAD models. 
These models are desirable for acoustic simulation because they have well-defined geometries and the most accurate semantic labels. 
Therefore, we use CAD models from the 3D-FRONT dataset~\cite{fu20213d}, which contains 18,968 diversely furnished rooms in 6,813 irregularly shaped houses/scenes. 
These different types of rooms (e.g., bedrooms, living rooms, dining rooms, and study rooms) are diversely furnished with varying numbers of furniture objects in meaningful locations. 
This differs from prior methods that use empty shoebox-shaped rooms~\cite{grondin2020bird,ko2017study}, because room shapes and the existence of furniture will significantly modify the acoustic signature of the room, including shifting the room modes in the low frequency. 
3D-FRONT dataset is designed to have realistic scene layouts, and has received higher human ratings in subjective studies~\cite{fu20213d}. 
Generating audio data from these models allows us to better approximate real-world acoustics. 

\subsection{Semantic Acoustic Material Assignment}
\label{sec:semantic-material}


Because the 3D-FRONT dataset also provides object semantics (i.e., object material labels), it is possible to assign more meaningful acoustic materials to individual surfaces or objects in the scene. 
For example, an object with ``window'' description is likely to be matched with several types of window glass material from the acoustic material database. 
\emph{SoundSpaces} dataset~\cite{chen2020soundspaces} also utilizes scene labels by using empirical manual material assignment (e.g., acoustic materials of carpet, gypsum board, and acoustic tile are assumed for floor, wall, and ceiling classes), creating a one-to-one visual-acoustic material mapping for the entire dataset. 
This approach works for a small set of known material types. 
Instead, we present a general and fully automatic method that works for unknown materials with unstructured text descriptions. 

To start with, we retrieve measured frequency-dependent acoustic absorption coefficients for $2,042$ unique materials from a room acoustic database~\cite{kling_2018}. 
The descriptions of these materials do not directly match the semantic labels of objects in the 3D-FRONT dataset. 
Therefore, we present a method to calculate the semantic similarity between each label and material description using natural language processing (NLP) techniques. 
In NLP research, sentences can be encoded into fixed length numeric vectors known as sentence embedding~\cite{mishra2019survey}. 
One goal of sentence embedding is to find semantic similarities to identify text with similar meanings. 
Transformer networks have been very successful in generating good sentence embeddings~\cite{liu2020survey} such that sentences with similar meanings will be relatively close in the embedding vector space. 
We leverage a state-of-the-art sentence transformer model~\footnote{https://huggingface.co/sentence-transformers/distiluse-base-multilingual-cased-v2}~\cite{reimers2019sentence} that calculates an embedding of dimension $512$ for each sentence. 
Our assignment process is described in Algorithm~\ref{alg:matching}. 
    \begin{algorithm}[h] 
    \caption{Semantic material assignment}\label{alg:matching}
    \KwInput{Object name string $s_0$}
    \KwData{Embedding model $f:s\longmapsto \mathbb{R}^{512}$; \\
    Acoustic material names $\{m_1,m_2,...,m_N\}$}
    \KwOutput{Acoustic material assignment}
    $e_0 \gets f(s_0)$ \tcp*{embedding for object name}
    \For{$i\gets 1$ \KwTo $N$ }{
        $e_i \gets f(m_i)$ \tcp*{embedding for acoustic material}
        $w_i \gets max\{0, cosineSimilarity(e_0, e_i)\}$\;
    }
    Assign $m_i$ with probability $P(X=m_i)=w_i/\Sigma_{i=1}^{N} w_i$\;
    \end{algorithm}

The key idea of our assignment is to use the cosine similarity (truncated to be non-negative) between the embedding vectors of the input and target material names as sampling weights to assign materials. 
Note that we do not directly pick the material with the highest score because for the same type of material, there are still different versions with different absorption coefficients (e.g., in terms of thickness, brand, painting). 
These slightly different descriptions of the same material are likely to have similar semantic distance to the 3D-FRONT material label being considered. Therefore, we use a probabilistic assignment to increase the diversity of our materials.

\subsection{Geometric-Wave Hybrid Simulation}
It is well known that geometric acoustic (GA) methods do not model low-frequency acoustic effects well due to the linear ray assumption~\cite{funkhouser1998,schissler2014high}. 
Therefore, we use a hybrid propagation algorithm that combines wave-based methods with GA. These wave-based methods can accurately model low-frequency wave effects, but their running time increases as the third or fourth power of the highest simulation frequency~\cite{raghuvanshi2009efficient}. Given the high time complexity of wave-based methods, we also want to use methods that are: (1) highly parallelizable so that dataset creation takes acceptable time on high-performance computing clusters; (2) compatible with arbitrary geometric mesh representations and acoustic material inputs; and (3) open-source so that the simulation pipeline can be reused by the research community. 
In this paper, we develop our hybrid simulation pipeline from a CPU-based GA software \emph{pygsound}~\footnote{https://github.com/GAMMA-UMD/pygsound} and a GPU-based wave FDTD software \emph{PFFDTD}~\cite{hamilton2021pffdtd}. 

\subsubsection{Inputs}
The scene CAD models from the 3D-FRONT dataset, each corresponding to several rooms with open doors, are represented in a triangle mesh format. 
Most GA methods have native support for 3D mesh input. The meshes are converted to voxels to be used as geometry input to the wave-based solver.
We sample source and receiver locations in each scene by grid sampling in all three dimensions with $1m$ spacing. This sampling yields from tens to thousands source and receiver pairs in each scene depending on its size. We perform collision checking to ensure all sampled locations have at least $0.2m$ clearance to any object in the scene. 

We assign acoustic absorption coefficients  according to the scheme presented in \S~\ref{sec:semantic-material}. 
These coefficients can be directly used by the GA method and integrated with the passive boundary impedance model used by the wave FDTD method~\cite{bilbao2015finite}. 
The GA method  also requires scattering coefficients, which account for the energy ratio between specular and diffuse reflections. 
Such data is less conventionally measured and is not available from the material database in \S~\ref{sec:semantic-material}. 
It is known that scattering coefficients tend to be negligible (e.g., $\leq 0.05$) for low-frequency bands~\cite{cox2006tutorial} handled by the wave method. 
Therefore, we sample scattering coefficients by fitting a normal distribution to $3$7 sets of  frequency-dependent scattering coefficients obtained from the benchmark data in \S~\ref{sec:benchmark}, which are only used by the GA method.

\subsubsection{Setup}
For the GA method, we set $20,000$ rays and $200$ maximum depth for specular and diffuse reflections. 
The GA simulation is intended for human aural range, while most absorption coefficient data is only valid for octave bands from 63Hz to 8,000Hz. 
The ray-tracing stops when the maximum depth is reached or the energy is below the hearing threshold. 

For the wave-based FDTD method, we set the maximum simulation frequency to 1,400Hz. 
The grid spacing is set according to $10.5$ points per wavelength. 
Our simulation duration is $1.5s$, which is sufficient to capture most indoor reverberation times and can be extended to generate additional data. 

\subsubsection{Automatic Calibration}
Before combining simulated IRs from two methods, one important step is to properly calibrate their relative energies. 
\citet{southern2011spatial} describe two objective calibration methods: (1) pre-defining a crossover frequency range near the highest frequency of the wave method and aligning the sound level of the two methods in that range; (2) calibrating the peak level from time-windowed, bandwidth-matched regions in the wave and the GA methods. 
Both calibration methods are used case-by-case for each pair of IRs. 
However, the first method is not physically correct, and the second method can be vulnerable when the direct sound is not known, as with occluded direct rays in the GA method. 
\citet{southern2013room} improved the second method by calculating calibration parameters once in free-field condition using a band-limited source signal. 

We use a similar calibration procedure. 
The calibration source and receivers have a fixed distance $r=1m$ in a large volume with absorbing boundary conditions, and the $90$ calibration receivers span a $90^{\circ}$ arc to account for the influence of propagation direction along FDTD grids. 
The source impulse signal is low-pass filtered at a cut-off frequency of $255$Hz.
Since the source signal is a unit impulse, the filtered signal is essentially the coefficients of the low-pass filter. 
The simulated band-limited IRs are truncated at twice the theoretical direct response time to further prevent any unwanted reflected wave. 
The calibration parameter for wave-based FDTD is computed as $\eta_w = \sqrt{E_s/E_r}$,
where $E_s$ is the total energy of the band-limited point source, and $E_r$ is the total energy at the receiver point. 
For multiple receiver points, $\eta_w$ takes the average value. 
During wave-based FDTD calibration, each received signal is multiplied by $\eta_w$, and we can calculate the difference between the calibrated signal and the band-limited source signal. 
As a result, we obtain a very low mean error of $0.50dB$ and a max error of $0.85dB$ among all calibration receivers. 

\begin{figure}[htbp]
  \includegraphics[width=\linewidth]{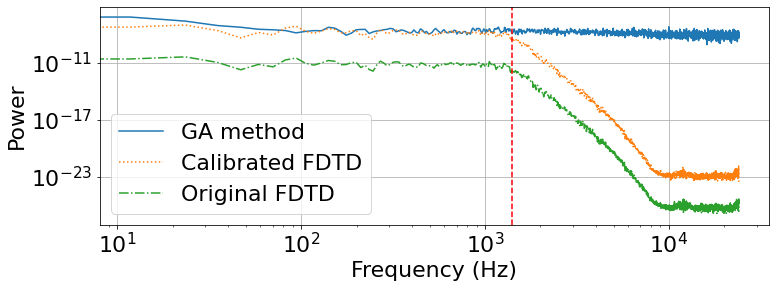}
  \caption{Power spectrum comparison between the original wave FDTD simulated IR and the calibrated IR. The vertical dashed line indicates the highest valid  frequency of the FDTD method. Our automatic calibration method ensures that  the GA and wave-based methods have consistent energy levels so that they can generate high quality IRs and plausible/smooth sound effects. 
  }
  \label{fig:calibration}
\end{figure}

For the GA method, we follow the same procedure though the process is simpler since the direct sound energy is explicit in most GA algorithms (i.e., $\frac{1}{r}$ scaled by some constant). 
Another calibration parameter $\eta_g$ is similarly obtained for the GA method. 
This calibration process ensures that the full-band transmitted energy from both methods will be $E=1$ at a distance of $1m$ from a sound source, although the absolute energy does not matter and the two parameters can be combined into one  (i.e., only use $\eta_w'=\eta_w/\eta_g$ for wave calibration). 
Figure~\ref{fig:calibration} shows an example of simulation results with and without calibration. 
Without properly calibrating the energies, there will be abrupt sound level changes in the frequency domain, which can create unnatural sound. 

\subsubsection{Hybrid Combination}
Ideally we would want to use the wave-based method for the highest possible simulation frequency. Besides the running time, one issue with FDTD scheme is the rising dispersion error with the frequency~\cite{lehtinen2003time}. 
As a remedy, the FDTD results are first high-pass filtered at a very low frequency (e.g., 10Hz) to remove some DC offset and then low-pass filtered at the crossover frequency to be combined with GA results. 
We use a Linkwitz-Riley crossover filter~\cite{linkwitz1976active} to avoid ringing artifacts near the crossover frequency, harnessing its use of cascading Butterworth filters. 
The crossover frequency in this work is chosen to be $1,400$Hz to fully utilize the accuracy of wave simulation results. 
Higher simulation crossover frequencies could be used at the cost of increased FDTD simulation time. 

\subsection{Analysis and Statistics}

\paragraph{Runtime}
The runtime of our hybrid simulator depends on specific computational hardware. 
We utilize a high-performance computing cluster with 20 Intel Ivy Bridge E5-2680v2 CPUs and 2 Nvidia Tesla K20m GPUs on each node. 
On a single node, our simulator requires about $4,000$ computing hours for the wave-based FDTD method and about $2,000$ computing hours for the GA method to generate all data. 

\begin{figure}
     \centering
     \begin{subfigure}[b]{\linewidth}
         \centering
         \includegraphics[width=\linewidth]{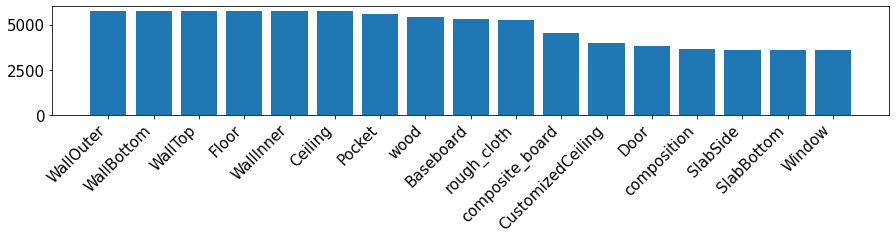}
         \caption{Occurrence of top visual material names.}
     \end{subfigure}
     \hfill
     \begin{subfigure}[b]{\linewidth}
         \centering
         \includegraphics[width=\linewidth]{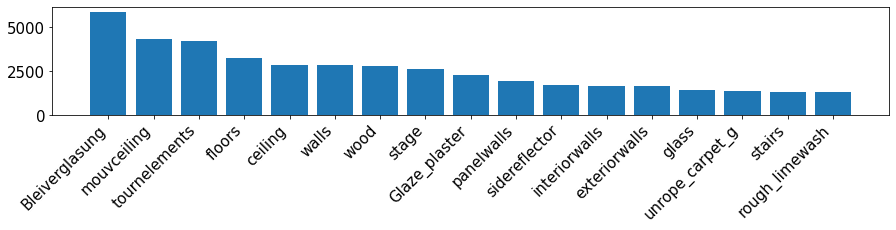}
         \caption{Occurrence of top acoustic material names.}
     \end{subfigure}

    \caption{We highlight the most frequently used materials in our approach for generating the IR dataset. The acoustic database also contains non-English words, which are handled by a pre-trained multi-lingual language model. }
    \label{fig:mat_stat}
\end{figure}

\paragraph{Distributions}
More than 5,000 scene/house models are used. 
On average, each scene uses 22.5 different acoustic materials. 
We assign $1,955$ unique acoustic materials (out of $2,042$) from the material database, and the most frequently used materials are several versions of brick, concrete, glass, wood, and plaster. 
The occurrence of most frequently used materials are visualized in Figure~\ref{fig:mat_stat}. 

\begin{figure}[htbp]
  \includegraphics[width=\linewidth]{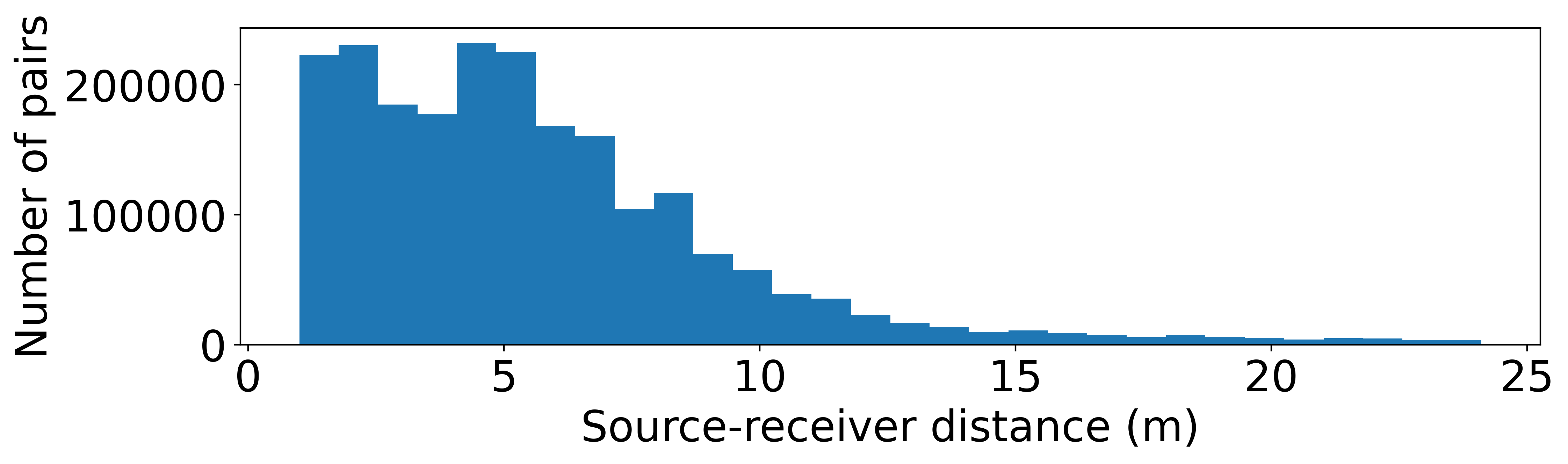}
  \caption{Distance distribution between source and receiver pairs in our scene database. 3D grid sampling is used in each scene followed by collision detection between sampled locations and objects in the scene. The IRs vary based on relative positions of the source and the receiver.
  }
  \label{fig:dist-stat}
\end{figure}

\begin{figure}[htbp]
  \includegraphics[width=\linewidth]{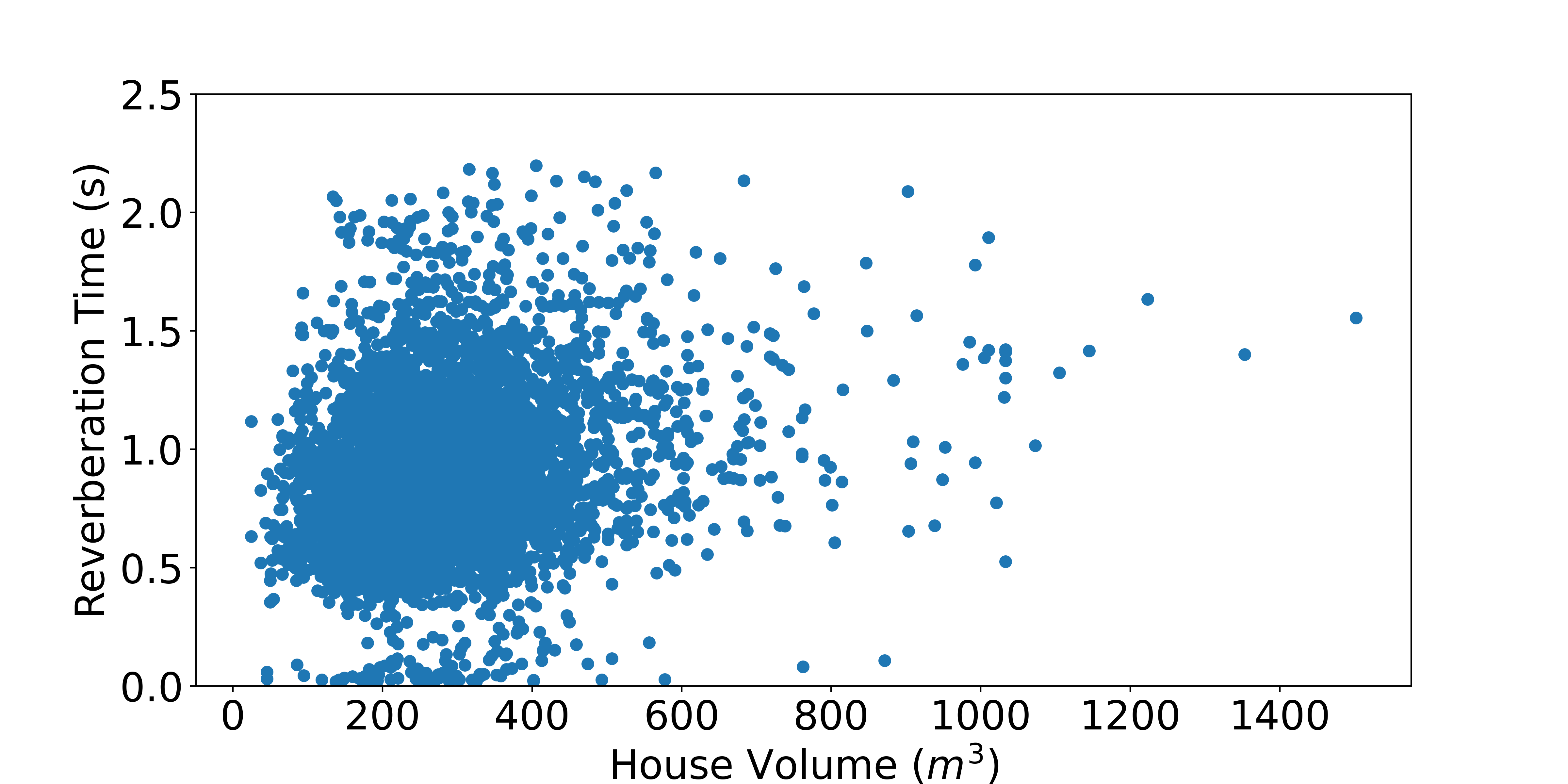}
  \caption{Statistics of house/scene volumes and reverberation times. We see a large variation in reverberation times, which is important for speech processing and other applications.
  }
  \label{fig:house-stat}
\end{figure}

The distribution of distances between all source and receiver pairs are visualized in Figure~\ref{fig:dist-stat}. 
We also show the relationship between the volume of each 3D house model and the reverberation time for that model in Figure~\ref{fig:house-stat} to highlight the wide distribution of our dataset. 
Overall, we have a balanced distribution of the reverberation times in the normal range. 

%% file: 4-evaluation.tex
\section{Acoustic Evaluation}

In this section, we evaluate the accuracy of our IR generation hybrid algorithm. We use a set of real-world acoustic scenes that have measured IR data to evaluate the effectiveness and accuracy of our hybrid simulation method. 

\subsection{Benchmarks}
\label{sec:benchmark}

Several real-world benchmarks have been proposed to investigate the accuracy of acoustic simulation techniques. 
A series of three round-robin studies~\cite{vorliander1995international,bork2000comparison,bork2005report1,bork2005report2} have been conducted on several acoustic simulation software systems by providing the same input and then comparing the different simulation results with the measured data. 
In general, these studies provide the room and material descriptions as well as microphone and loudspeaker specifications including locations and directivity. 
However, the level of detailed characteristics, in terms of complete 3D models and  consistent measured acoustic material properties tend to vary. 
Previous round-robin studies have identified many issues (e.g., uncertainty in boundary condition definitions) in terms of simulation input definitions for many simulation packages, which can result in poor agreement between simulation results and real-world measurements. 
A more recent benchmark, the BRAS benchmark~\cite{aspock2020benchmark}, contains the most complete scene description and has a wide range of recording scenarios.
We use the BRAS benchmark to evaluate our simulation method. 
Three reference scenes (RS5-7) are designed as diffraction benchmarks and we use them to evaluate the performance of our hybrid simulator, especially at lower frequencies.

The 3D models of the reference scenes along with frequency-dependent acoustic absorption and scattering coefficients are directly used for our hybrid simulator.  
We use these three scenes because they are considered difficult for the geometric method alone~\cite{brinkmann2019round}. 

\subsection{Results}

\begin{figure}
     \centering
     \begin{subfigure}[b]{\linewidth}
         \centering
         \includegraphics[width=\linewidth]{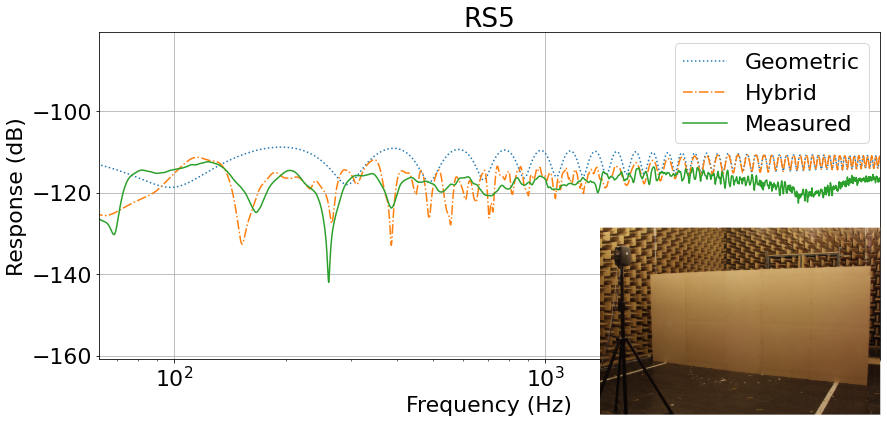}
         \caption{RS5: simple diffraction with infinite edge.}
     \end{subfigure}
     \hfill
     \begin{subfigure}[b]{\linewidth}
         \centering
         \includegraphics[width=\linewidth]{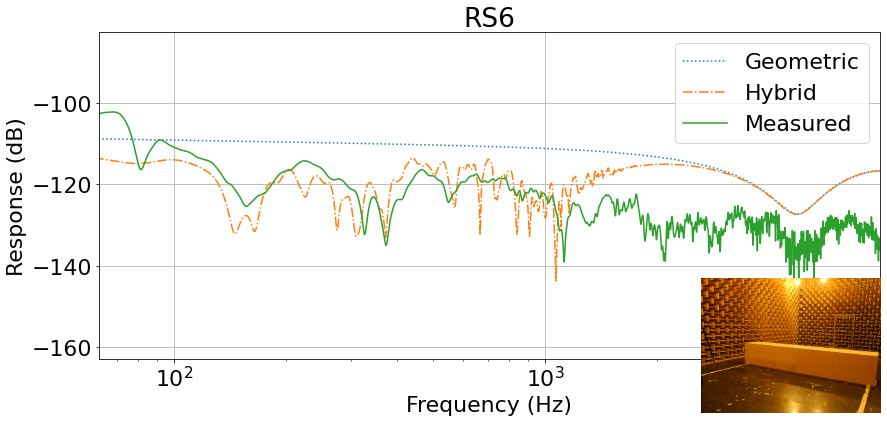}
         \caption{RS6: diffraction with infinite body.}
     \end{subfigure}
     \hfill
     \begin{subfigure}[b]{\linewidth}
         \centering
         \includegraphics[width=\linewidth]{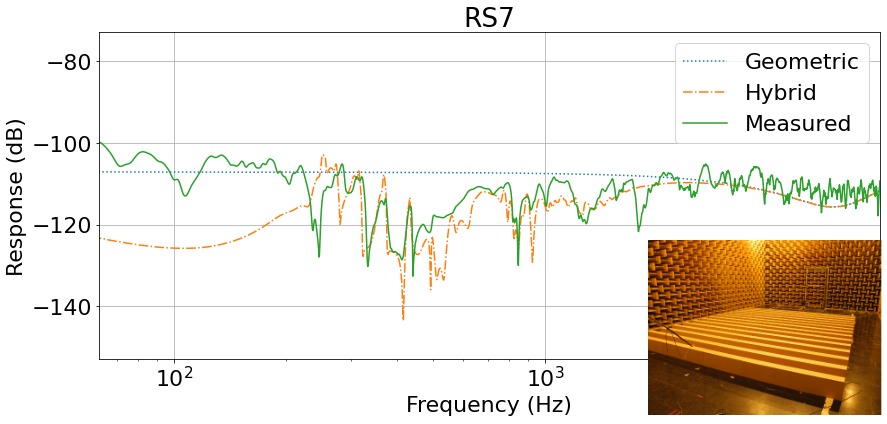}
         \caption{RS7: multiple diffraction (seat dip effect)}
     \end{subfigure}
        \caption{Frequency responses of geometric and hybrid simulations compared with measured IRs in BRAS benchmarks RS5-7~\cite{aspock2020benchmark}. Images of each setup are attached in the corners of the graph. We notice that the IRs generated using our hybrid method closely match with the measure IRs, as compared to those generated using GA methods. This demonstrates the higher quality and accuracy of our IRs as compared to the ones generated by prior GA methods highlighted in Table 1.}
        \label{fig:freq_response}
\end{figure}

We use the room geometry, source-listener locations, and material definitions as an input to our simulation pipeline.  Note that the benchmark only provide absorption and scattering coefficients, and no impedance data is directly available for wave solvers. 
Thus, we only use fitted values rather than exact values. 
The IRs generated by the GA method and our hybrid method and the measured IRs from the benchmark are compared in the frequency domain in Figure~\ref{fig:freq_response}. 
In these scenes, the source and receiver are placed on different sides of the obstacle and the semi-anechoic room only has floor reflections. 
In the high frequency range, there are fewer variations in the measured response, and both methods capture the general trend of energy decay despite response levels not being perfectly matched. 
This demonstrates that our hybrid sound simulation pipeline is able to generate more accurate results than the GA method for complex real-world scenes.

%% file: 5-application.tex
\section{Applications}

We use our dataset on three speech processing applications that use deep learning methods. Synthetic IRs have been widely used for training neural networks for automatic speech recognition, speech enhancement, and source separation (as described in \S~\ref{sec:intro}). We evaluate the benefits of generating a diverse and high-quality IRs dataset over prior methods used to generate synthetic IRs.

In general, these speech tasks use IR datasets to augment anechoice speech data to create synthetic distant training data, whereas the test data is reverberant data recorded in the real world. 
When high-quality IR datasets are used, the training set is expected to generalize better on the test data. 
We simulate distant speech data $x_d[t]$ by convolving anechoic speech $x_{c}[t]$ with different IRs $r[t]$ and adding environmental noise $n[t]$ from BUT ReverbDB \cite{szoke2019building} dataset using

{\belowdisplayskip 0.4\belowdisplayshortskip
\begin{equation}\label{reverberant_speech}
\begin{aligned}[b]
    x_{d}[t] = x_{c}[t] \circledast r[t] + n[t + l].
\end{aligned}
\end{equation}
}
Then the data is used by different training procedure and neural network architectures on respective benchmarks. 
In the following tests, we also examine variations of our datasets: \textbf{GA} (geometric method only), \textbf{FDTD} (only up to 1,400Hz), and \textbf{GWA} (hybrid method).

\subsection{Automated Speech Recognition}
\label{sec:asr}


\begin{table}[ht]
\caption{\label{table_ami}Far-field ASR results obtained for the AMI corpus.}
\centering
{\begin{tabular}{lc}
\hline
\textbf{IR used}      & \textbf{WER[\%]$\downarrow$} \\
\hline
None (anechoic) & 64.2 \\
BIRD~\cite{grondin2020bird}+\citet{ko2017study} & 48.4 \\
BUT Reverb~\cite{szoke2019building} & \textbf{46.6} \\
\hline
SoundSpaces (manual)~\cite{chen2020soundspaces} & 51.2 \\
SoundSpaces-semantic & \textbf{50.3} \\
\hline
FDTD & 58.1 \\
GA & 48.1 \\
GWA subset & 48.5 \\
\textbf{GWA (ours)}  & \textbf{47.7}\\
\hline
\end{tabular}}
\end{table}

Automatic speech recognition (ASR) aims to convert speech data to text transcriptions. 
The performance of ASR models is measured by the \emph{word error rate (WER)}, which is the percentage of incorrectly transcribed words in the test data. 
The AMI speech corpus~\cite{ami} consisting of $100$ hours of meeting recording is used as our benchmark. 
And we use the \emph{Kaldi} toolkit~\cite{povey2011kaldi} to run experiments on this benchmark. 
We randomly select $30,000$ IRs out of 2M synthetic IRs in GWA to augment the anechoice training set in AMI, and report the WER on the real-world test set. 
A lower WER indicates that the synthetic distant speech data used for training is closer to real-world distant speech data. We highlight the improved accuracy obtained using GWA over prior synthetic IR generators in Table \ref{table_ami}. 

In this benchmark, we include more comparisons with prior datasets from Table~\ref{tab:comparison} and we also incorporate SoundSpaces \cite{Chen_2018_ECCV} data into our pipeline. We have replaced the original manual material assignment in SoundSpaces with our semantic assignment (\textbf{SoundSpaces-semantic}), and observe a $0.9\%$ WER improvement. To study the effect of model diversity, we use a subset of house models from 3D-FRONT (\textbf{GWA subset}) that has the same number (103) of scenes as SoundSpaces, this results in a $0.8\%$ degradation in WER compared with our full dataset. Overall, we see that our \textbf{GWA} dataset outperforms prior synthetic datasets, where \textbf{BIRD+\citet{ko2017study}} uses an image-method simulator in shoebox-shaped rooms and \textbf{SoundSpaces} uses a ray-tracing based simulator, and has the closest WER to that of the real-world \textbf{BUT Reverb} \cite{szoke2019building} dataset, due to both the inclusion of more diverse 3D scenes and our semantic material assignment.


\subsection{Speech Dereverberation}

Speech dereverberation aims at converting a reverberant speech signal back to its anechoic version to enhance its intelligibility. 
For our tests, we use SkipConvNet \cite{skipconvnet}, a U-Net based speech dereverberation model.
The model is trained on the 100-hour subset of Librispeech dataset \cite{libri}. 
The reverberant input to the model is generated by convolving the clean Librispeech data with our synthetically generated IRs. 
\begin{table}[ht]
\caption{\label{tab:SE-SRMR} Speech enhancement results trained using different synthetic IR datasets. }
\centering
\begin{tabular}{lc}
\hline
\textbf{IR used}      & \textbf{SRMR$\uparrow$} \\
\hline
None (anechoic) & 4.96 \\
SoundSpaces & 7.44 \\
GA         & 6.01 \\
FDTD        & 4.78 \\
\textbf{GWA (ours)}      & \textbf{8.14} \\
\hline
\end{tabular}
\end{table}
We test the performance of the model on real-world recordings from the VOiCES dataset \cite{richey2018voices} which contain reverberant Librispeech data. 
We report the \emph{speech-to-reverberation modulation energy ratio (SRMR)} \cite{falk2010non} over the test set. 
A higher value of SRMR indicates lower reverberation and higher speech quality. As seen from Table~\ref{tab:SE-SRMR}, our proposed dataset obtains better dereverberation performance as compared to all other datasets. 


\subsection{Speech Separation}

We train a model to separate reverberant mixtures of two speech signals into its constituent reverberant sources. We use the Asteroid~\cite{Pariente2020Asteroid} implementation of the DPRNN-TasNet model \cite{luo2020dualpath} for our benchmarks. The 100-hour split of the Libri2Mix \cite{cosentino2020librimix} dataset is used for training. 
We test the model on reverberant mixtures generated from the VOiCES dataset. We report the improvement in \emph{scale-invariant signal-to-distortion ratio (SI-SDRi)} \cite{sisdri} to measure separation performance. Higher SI-SDRi implies better separation. As seen from Table~\ref{tab:SS}, our GWA outperforms alternative approaches.

\begin{table}[ht]
\caption{\label{tab:SS} SI-SDRi values reported for different IR generation methods. We report results separately for the four rooms used to capture the test set (higher is better).}
\begin{tabular}{lcccc}
\hline
\multirow{2}{*}{\textbf{IR used}} & \multicolumn{4}{c}{\textbf{SI-SDRi$\uparrow$}}                                                                                       \\
\cline{2-5}
                        & \multicolumn{1}{l}{\textbf{Room 1}} & \multicolumn{1}{l}{\textbf{Room 2}} & \multicolumn{1}{l}{\textbf{Room 3}} & \multicolumn{1}{l}{\textbf{Room 4}} \\
\hline
SoundSpaces                     & 1.85                       & 1.74                       & 1.02                       & 1.80      \\
GA                     & 2.25                       & 2.55                       & 1.44                       & 2.55                       \\
FDTD                    & 2.36                       & 2.43                       & 1.33                       & 2.46                       \\
\textbf{GWA (ours)}                  & \textbf{2.94}                       & \textbf{2.76}                       & \textbf{1.86}                       & \textbf{2.91} \\
\hline 
\end{tabular}
\end{table}

%% file: 6-conclusion.tex
\section{Conclusion and Future Work}

We introduced a large new audio dataset of synthetic room impulse responses  and the simulation pipeline, which can take different scene configurations and generate higher quality IRs. 
We demonstrated the improved  accuracy of our hybrid geometric-wave simulator on three difficult scenes from the BRAS benchmark.
As compared to prior datasets, GWA has more scene diversity  than recorded datasets, and has more physically accurate IRs than other synthetic datasets. We also use our dataset with audio deep learning to improve the performance of speech processing applications.

Our dataset only consists of synthetic scenes, and may not be as accurate as real-world captured IRs. In many applications, it is also important to model ambient noise. 
In the future, we will continue growing the dataset by including more 3D scenes to further expand the acoustic diversity of the dataset. We plan to evaluate the performance of other audio deep learning applications.